\documentclass{pnastwo}

\usepackage{graphicx}
\usepackage{amssymb}
\usepackage{amssymb,amsfonts,amsmath}

\url{www.pnas.org/cgi/doi/10.1073/pnas.1208875109}
\copyrightyear{2012}
\issuedate{July 24, 2012}
\volume{vol. 109}
\issuenumber{no. 30}
\begin{document}

\title{Helioseismology challenges models of solar convection}

\author{Laurent Gizon\affil{1}{Max-Planck-Institut f\"{u}r Sonnensystemforschung, Max-Planck-Stra{\ss}e 2, 37191 Katlenburg-Lindau, Germany}\affil{2}{Institut f\"{u}r Astrophysik, Georg-August-Universit\"{a}t G\"{o}ttingen, Friedrich-Hund-Platz 1, 37077 G\"{o}ttingen, Germany}
\and Aaron C. Birch\affil{1}{}}

\contributor{Submitted to Proceedings of the National Academy of Sciences of the United States of America}

\maketitle

\begin{article}



\dropcap{C}onvection is the mechanism by which energy is transported through the outermost $30\%$ of the Sun \cite{Gilman2000}. Solar turbulent convection is notoriously difficult to model across the entire convection zone where the density spans many orders of magnitude. In this issue of PNAS, Hanasoge et al. \cite{Hanasoge2012} employ recent helioseismic observations to derive stringent empirical constraints on the amplitude of large-scale convective velocities in the solar interior. They report an upper limit that is far smaller than predicted by a popular hydrodynamic numerical simulation.

Historically, great advances in our understanding of the solar interior have been due to helioseismology, the study of five-minute solar internal oscillations \cite{JCD2002}. In the mid 1980s global-mode frequencies were used to measure the depth of the solar convective envelope at $0.71$ solar radius, 
deeper than previous expectations based on underestimated opacities. Another spectacular achievement was the inference of solar rotation as a function of radius and latitude. 
The bulk of the convective envelope rotates differentially, faster at the equator than at high latitudes. At the base of the convection zone is a zone of rotational shear, known as the tachocline, which now plays a central role in theories of the solar dynamo \cite{Charbonneau2010}. Despite valuable attempts, none of the above solar features were confidently predicted by models. Whenever helioseismology opens a new window into the solar interior, surprises are possible.

The work of Hanasoge et al. \cite{Hanasoge2012} is perhaps the most notable helioseismology result since the launch of the Helioseismic and Magnetic Imager (HMI) \cite{Scherrer2012} onboard NASA's Solar Dynamics Observatory (SDO). 
HMI measures the motions on the solar surface caused by the random superposition of seismic waves excited by near-surface convection. Full-Sun Doppler velocity images are captured every 45 seconds by a 16 million pixel camera. Hanasoge et al. used this unique combination of high resolution and full spatial coverage to carry out high-precision helioseismology of large-scale solar convection.

To first order, convective flows do not affect the global-mode frequencies of solar oscillations. On the other hand, in time-distance helioseismology \cite{Duvall1993}, wave travel times are linearly sensitive to subsurface flows \cite{Gizon2010}. Time-distance helioseismology uses spatio-temporal correlations of the random wave field to measure the travel times of solar acoustic waves between distant locations on the solar surface. The two-point correlation function contains fundamental information: it plays a role similar to a Green's function. One application of time-distance helioseismology has been the study of solar supergranulation\cite{Duvall2000}; This preferred scale of convection with typical length scale $35$~Mm has remained a challenge for theory since its discovery fifty years ago \cite{RieutordRincon2010}.

\begin{figure}
\begin{center}
\includegraphics[width=0.5\linewidth]{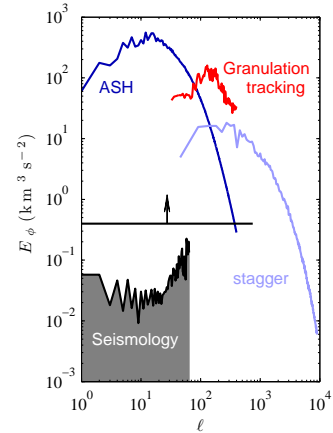}
\end{center}
\caption{Comparison of kinetic energy spectra $E_\phi$ of longitudinal solar velocities versus spherical harmonic degree $\ell$. The black curve above the gray area shows the observational upper limit from SDO/HMI helioseismology at radius $0.96 R_\odot$ and one day averaging \cite{Hanasoge2012}. The red curve is from surface velocities measured by tracking granules using SDO/HMI intensity images \cite{Roudier2012}. Notice the excess power at $\ell\sim 120$ due to supergranulation. Results from snapshots of numerical simulations of convection sliced at $0.98 R_\odot$ are given by the blue and light blue curves. The blue curve shows the spectrum from an ASH spherical-shell simulation \cite{Miesch2008} and the light blue curve shows the spectrum from a stagger \cite{Stein2006} near-surface radiative compressible simulation of size $96\times96\times20$~Mm$^3$. The horizontal black line and associated arrow shows a theoretical lower limit based on global dynamics arguments \cite{Miesch2012}, assuming mode equipartition over $\ell<750$. We defined $E_\phi$ at radius $r$ such that $\langle v_\phi^2 \rangle /2 = \sum_{\ell\ge 0} E_\phi(\ell)/r$, where $v_\phi$ is the longitudinal (prograde) component of the velocity field and the expectation value is approximated by a horizontal average (consistent with \cite{Rieutord2010} in the limit of large $\ell$).}
\end{figure}

Hanasoge et al. \cite{Hanasoge2012} measured east-west travel-time differences using a deep-focusing, quadrant geometry, for a range of integration times, $T$, up to $96$~hr. Travel times are dominated by stochastic noise due to the incoherent superposition of solar seismic waves. According to Hanasoge et al., an upper limit on convective velocities at harmonic degree $\ell$ is $v_{\phi, \ell} < \epsilon_{\rm sg} \, \tau_\ell / C_\ell$. In this expression, $\epsilon_{\rm sg}\ll 1$ is the signal-to-noise ratio of the travel times (dominated by the contribution from supergranulation) and $\tau_\ell$ is the rms travel time at scale $\ell$ only (dominated by stochastic noise). The quantity $\epsilon_{\rm sg}$ may be estimated from the $T$-dependence of the variance of the travel times, since the variance of pure noise behaves like $1/T$ \cite{GB2004}. The measurement of $\epsilon_{\rm sg}$ by Hanasoge et al. \cite{Hanasoge2012} is an advance compared to the earlier approach of Hanasoge et al. \cite{Hanasoge2010} where $\epsilon_{\rm sg}=1$ was assumed. The calibration constant $C_\ell$ used above to convert travel times into velocities (see \cite{Hanasoge2012}, Figure 5 of supplementary material) is the result of numerical forward modelling and leads to an upper limit on the velocity at the target depth.

The interpretation of travel-time measurements is a topic of current research, especially regarding the effect of time-dependent turbulent velocities. Hanasoge et al. assumed frozen convection to obtain the calibration constant $C_\ell$, which is only an approximation when the lifetime of convection is less than $T$. This may lead to an underestimation of the solar velocities. Also the vertical correlations of the convective velocities were ignored: Eddies of vertical sizes less than the first Fresnel zone cannot be detected. These two points may potentially affect the inferred upper limit on the convective velocities. Furthermore, experience with other experiments suggests that systematic effects, e.g. center to limb effects, could also play a role at the m/s level.

It is enlightening to consider the helioseismology results in the context of existing models of convection. Hanasoge et al. \cite{Hanasoge2012} showed that in the range $\ell<60$ their helioseismology upper limit for longitudinal convective velocities at radius $0.92 R_\odot$ is orders of magnitude less than what is predicted by an Anelastic Spherical Harmonic (ASH) hydrodynamic simulation \cite{Miesch2008}. This is a main concern of the authors. For the outside commentator, there is no clear way to reconcile this severe disagreement.

Here we supplement the comparison with additional theoretical, numerical, and observational constraints, which we have combined in the figure. We chose the kinetic energy density $E_\phi(\ell)$ to characterize the strength of longitudinal convective velocities \cite{Rieutord2010}, which is a partial but useful description of convection.

Miesch et al. \cite{Miesch2012} recently obtained an interesting theoretical lower limit of $30$~m\,s$^{-1}$ for convective velocities at radius $0.95R_\odot$ for scales $\ell\lesssim 750$. This calculation is based on the idea that the observed large-scale flows (differential rotation and meridional circulation) are maintained by convective Reynolds stresses. Partitioning the kinetic energy evenly over all modes $\ell<750$, we find $E_\phi>0.4$~km$^3$\,s$^{-2}$, which is well above the helioseismology upper limits at the lowest $\ell$ values (see figure). More work is needed to determine the $\ell$ dependence of this theoretical lower limit.

The ASH simulation is truncated at radius $0.98 R_\odot$, above which additional physics is needed, for example compressibility and radiative transfer. Convection in the near-surface layers has been
modelled with great success (judging from comparisons
with surface observations \cite{Nordlund2009})
using fully compressible radiative simulations
in local Cartesian simulation boxes.
For example a recent stagger simulation \cite{Stein2006}
covers $r>0.97 R_\odot$, which overlaps
in radius with the ASH simulation.
The kinetic energy spectra of the two simulations at
$r=0.98 R_\odot$ roughly agree around $\ell\sim 150$.
This suggests that ASH is capturing some of the general dynamics
there, despite using simplified physics
and missing the driving by strong cooling at
the solar surface.

SDO/HMI provides at least two other means than helioseismology to observe surface flows. The first method is direct Doppler measurements of the line-of-sight component of velocity \cite{Hathaway2000}. The second method is based on tracking the motions of granules or other small features and provides both components of the horizontal velocities.
In the Figure, we plot the granulation-tracking result from Roudier et al. \cite{Roudier2012}, who employed SDO/HMI intensity images.
Notice that the ASH kinetic energies are above the granulation-tracking value for $\ell<80$. This is surprising since one would expect the convective velocities to decrease in amplitude with increasing depth.

A striking feature in the granulation-tracking curve is the excess kinetic energy at $\ell \sim 120$ due to supergranulation. Current simulations of convection are not ideal for modelling the supergranulation in detail; currently this scale is near the largest scale of the stagger code and the smallest scale of ASH. The helioseismology inferences from Hanasoge et al.
stop short of the supergranular scale; the method of analysis
was not optimized for that purpose. Overall, a better observational
coverage and theoretical understanding of the intermediate spatial scales
would help connect the local and global scales of convection.
The next generation of convection models are expected to cover the supergranulation range.

Assuming that the helioseismology upper limit on convective velocities from Hanasoge et al. \cite{Hanasoge2012} can be taken at face value, this will force a rethinking of the large-scale dynamics of the solar convective zone.
One particular question is how to model very highly turbulent regimes, e.g. by including deep thermal plumes \cite{Rieutord1995}.

Any viable theory of convection ought to explain convection in other stars. In this respect asteroseismology may play an important role. The observed amplitudes of oscillation in other Sun-like stars \cite{Michel2008} contain information about the vigour of surface convection in these stars, which in turn will place constraints on stellar convection models.

\begin{acknowledgments}
We thank Mark Miesch (Anelastic Spherical Harmonic code), Thierry Roudier (granulation tracking) and Robert Stein (stagger code) for making their data available. We have received support from Deutsches Zentrum für Luft- und Raumfahrt (DLR) grant ''German Data Center for SDO'', European Research Council Starting Grant ''Seismic Imaging of the Solar Interior'', and Deutsche Forschungsgemeinschaft SFB-963 grant ''Astrophysical Flow Instabilities and Turbulence.''
\end{acknowledgments}

\end{article}

\end{document}